\documentclass[14pt,oneside,english]{extbook}
\usepackage[T1]{fontenc}
\usepackage[latin9]{inputenc}
\usepackage{geometry}
\usepackage{color}
\usepackage{babel}
\usepackage{array}
\usepackage{units}
\usepackage{multirow}
\usepackage{amsmath}
\usepackage{graphicx}
\usepackage[unicode=true]
 {hyperref}

\makeatletter

\newcommand{\lyxmathsym}[1]{\ifmmode\begingroup\def\b@ld{bold}
  \text{\ifx\math@version\b@ld\bfseries\fi#1}\endgroup\else#1\fi}

\providecommand{\tabularnewline}{\\}

\date{}
\usepackage{breakurl, color, babel, textcomp,graphicx, setspace, lineno, lmodern, hyperref}
\hypersetup{
colorlinks=true,
linkcolor=red,
citecolor = blue,
urlcolor=blue,
pdfpagemode=FullScreen}

\usepackage{mathpazo}

\makeatother

\begin{document}
\title{\textbf{\Large{}ELECTROPOLYMERIZATION: STUDIES AND APPLICATIONS}}

\maketitle
\tableofcontents{}

\chapter[Electrodes and Double Layers]{{\Large{}ELECTROCHEMICAL SYSTEMS: ELECTRODES AND DOUBLE LAYERS}}
\begin{center}
{\large{}Asghar Aryanfar}\footnote{Corresponding Author: Email: \href{mailto:aryanfar\%40caltech.edu}{aryanfar@caltech.edu}}{\large{}$^{*,\dagger}$,
Agustin J. Colussi$^{*}$, Laleh M. Kasmaee$^{*}$, Michael R. Hoffmann$^{*}$}{\large\par}
\par\end{center}

\begin{center}
\emph{\small{}$*$ California Institute of Technology, 1200 E California
Blvd, Pasadena, CA 91125}{\small\par}
\par\end{center}

\begin{center}
\emph{\small{}$\dagger$ Bahçe\c{s}ehir University, 4 Ç\i ra\u{g}an
Cad, Be\c{s}ikta\c{s}, Istanbul, Turkey 34349}{\small\par}
\par\end{center}

\section{Abstract}

Electropolymerization plays a critical role in the electrochemical
systems. In this chapter, we address such role within the context
of interplay between kinetics and energetics. The trains of chin radical
reactions leads to the formation of thin films in electrochemical
devices. The structure of so-called solid electrolyte interphase (SEI)
during the initial charge/discharge cycles of the device of any kind
(i.e. rechargeable battery) on the surface of electrode directly controls
the the ultimate stability and longevity. In this chapter, we study
the morphological evolution of SEI, both in terms of transport and
thermodynamics within quantitative and qualitative contexts.

\section{Introduction}

Electropolymerization plays an important role in the operation of
rechargeable batteries in portable electronics and electric vehicles.
As an alkaline metal can react with the most organic solvents, a surface
film is formed during the initial charging/discharging processes.
This electrically insulating and ionically conductive interface is
named as the solid electrolyte interphase (SEI). \cite{Peled_79}

Electropolymerization typically occurs in the double layer region
(DL herein after). Also called as electrical double layer (EDL), such
structure appears on the surface of an object when it is exposed to
a fluid. The object might be a solid particle, a gas bubble, a liquid
droplet, or a porous body. The DL refers to two parallel layers of
charge surrounding the object. The first layer, the surface charge
(either positive or negative), consists of ions adsorbed onto the
object due to chemical interactions. The second layer is composed
of ions attracted to the surface charge via the Coulomb force, electrically
screening the first layer. This second layer is loosely associated
with the object. It is made of free ions that move in the fluid under
the influence of electric attraction and thermal motion rather than
being firmly anchored. It is thus called the \textquotedbl diffuse
layer\textquotedbl .

Interfacial DLs are most apparent in systems with a large surface
area to volume ratio, such as a colloid or porous bodies with particles
or pores (respectively) on the scale of micrometers to nanometers.
However, DLs are important to other phenomena, such as the electrochemical
behavior of electrodes. \cite{Block_78}

This layer and includes various organic and inorganic components.
On one hand, the formation of the SEI intrinsically consumes the anode
and electrolyte, leading to a low efficiency. Consequently, the SEI
effectively prevents the further physical contact between Li and the
solvent, therefore making Li dynamically stable in certain organic
electrolytes.\cite{Cheng_16} In particular, the SEI can adjust the
distribution of Li ions from the bulk electrolyte to the anode. This
layer is merely the result of competitive desolvation of ionic compounds
on the organic electrolytes. \cite{KASMAEE_16}

The SEI ultimately covers the Li electrodes in multilayer surface
films composed of organic or inorganic Li salts. Thereby, applying
an electrical field to Li electrodes enables electrochemical Li dissolution
and deposition to occur through these surface films. \cite{Aurbach_00}

\begin{figure}
\begin{centering}
\includegraphics[width=1\textwidth]{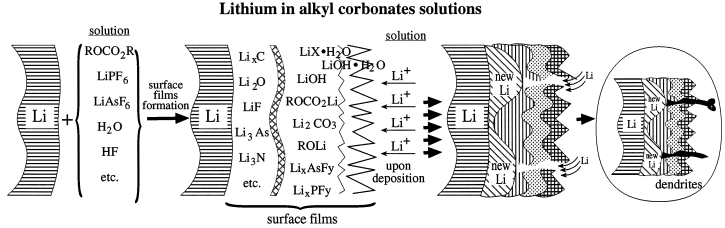}
\par\end{centering}
\caption{Various scenarios of SEI formation on lithium electrodes from organic
carbonates.\label{fig:Schematics}\cite{Aurbach_00}}
\end{figure}

The Figure \ref{fig:Schematics} shows that upon formation of SEI
layer, it interferes the morphology of deposition and therefore the
upcoming lithium ions cannot afford to perform uniform deposition.

Looking closer to the morphology, SEI is thin and fragile film and
stabilizes the redox reaction on the electrode surface. While this
film doesn't let bigger organic compounds to reduce further, it is
conductive to smaller charge carrier candidate ions such as lithium
($Li^{+}$) , Nickel ($Ni^{+}$), Magnesium ($Mg^{2+}$) or Zinc ($Zn^{+}$).

The morphology of SEI is highly effective on the rechargeable lithium
metal batteries as an optimal energy storage devices \cite{Parachin_16}.
$Li^{0}$ has an exceptional facility for growing dendrites, a feature
that causes battery degradation and ultimately failure \cite{ARYANFAR_14,ARYANFAR_15_2,Chen_15}.
Inhibition of such microstructures hinge on SEI generated from the
decomposition of most organic solvents at the negative potentials
required to reduce $Li^{+}$ \cite{Zhang_14}. The importance generally
ascribed to SEI is most objectively attested by recent reports which
emphasized that \textquoteleft . . .\emph{SEI formation is the most
crucial and least understood phenomena impacting battery technology}.
. .\textquoteright{} \cite{Soto_15}, and \textquoteleft . . .\emph{constructing
stable and efficient SEI is among the most effective strategies to
inhibit the dendrite growth and achieve superior cycling performance}.
. .\textquoteright{} \cite{An_16}. Previous attempts at improving
SEI properties have variously resorted to \textquoteleft . . .\emph{electrolyte
additives and surface modification of the cathode...(which) have been
shown to improve the formation of an effective SEI layer}. . .\textquoteright{}
and led to the conclusion that \textquoteleft . . .\emph{the formation
of the SEI depends largely on electrode materials, electrolyte salts,
and solvents involved}. . .\textquoteright{} \cite{Agubra_14}. 
\begin{table}
\begin{centering}
\begin{tabular}{|c|c|}
\hline 
\textbf{I} & $1M$$LiCLO_{4}$\tabularnewline
\hline 
\textbf{II} & $0.9M$ $LiClO_{4}$ + $0.1M$ $LiF$\tabularnewline
\hline 
\textbf{III} & $0.99M$ $LiClO_{4}$ + $0.01M$ $LiF$\tabularnewline
\hline 
\end{tabular}
\par\end{centering}
\caption{Electrolyte compositions. \label{tab:Elects}}
\end{table}

The composition and structure of SEI have also been intensively investigated
by diverse techniques, such as XPS, solid state NMR \cite{Blanc_13},
ellipsometry \cite{Lux_13}, sum-frequency generation spectroscopy
\cite{Yu_13}, electron microscopies \cite{LI_14_REVIEW}, neutron
scattering \cite{Bridges_12}, AFM \cite{LU_14}, electron paramagnetic
spectroscopy and matrix assisted laser desorption ionization (MALDI)
time of flight mass spectrometry \cite{XU_14}. Recent reviews, however,
have acknowledged that \textquoteleft ...\emph{many strategies have
been proposed to modify SEI structure. However, the modifying process
is still out of control in a bulk cell because the thickness, density
and ion conductivity cannot yet be rationally designed}\textquoteright{}
\cite{Cheng_16}.

\begin{figure}
\begin{centering}
\includegraphics[width=0.9\textwidth]{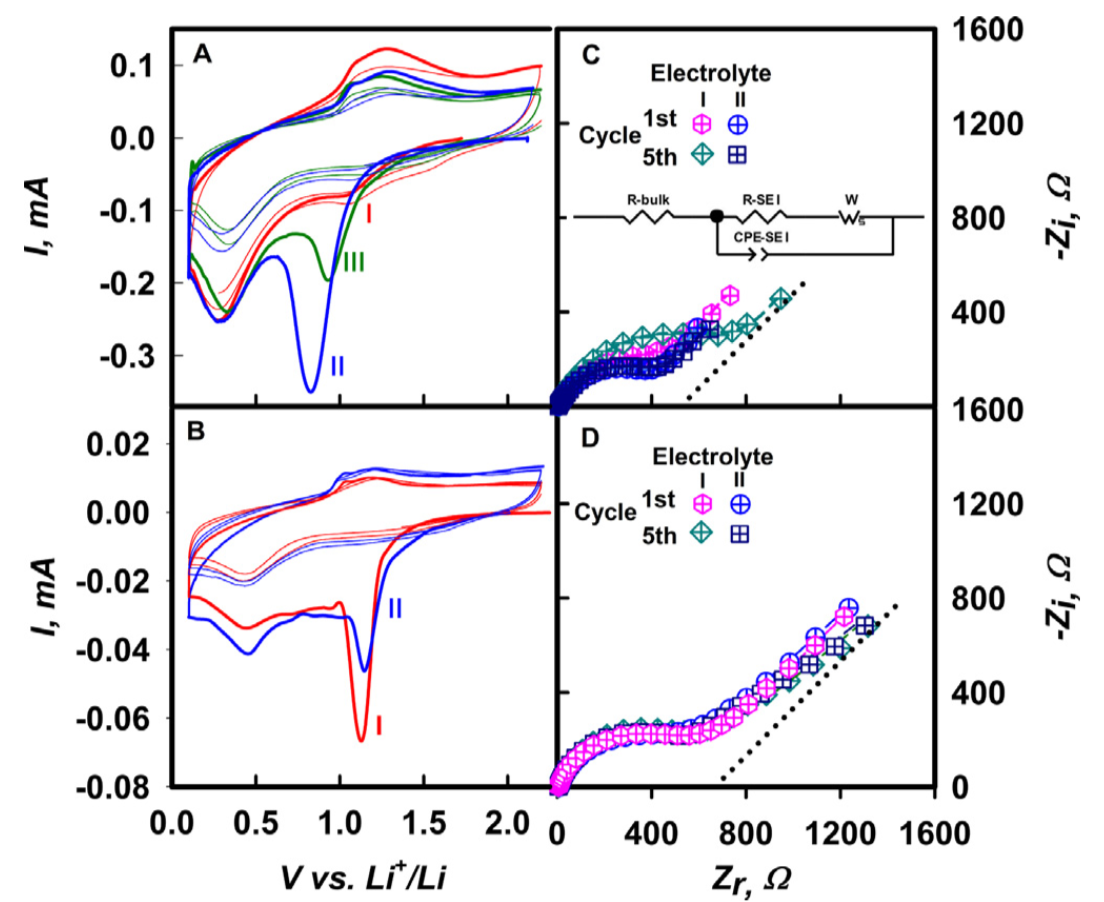}
\par\end{centering}
\caption{{\small{}CVs in Cu|electrolyte|Li cells filled with compositions of
Table \ref{tab:Elects}. Scan rates: }\textbf{\small{}A}{\small{}:$v=5.0\nicefrac{mV}{s}$
}\textbf{\small{}B:}{\small{}$v=0.5\nicefrac{mV}{s}$;}\textbf{\small{}
C,D: }{\small{}Corresponding OCV Nyquist diagrams for $V\in[OCV,0.1V]$
respectively; Inset: equivalent circuit. Dotted lines: Warbug's $n'=0.5$
slopes as a reference.\label{fig:CVs1}\cite{KASMAEE_16}}}
\end{figure}
One interpretation of this impasse is that SEI properties depend not
only on initial conditions, such as electrode materials, electrolyte
salts, solvents and additives, but on the procedure by which SEI are
generated. Thus, if the mechanisms of generation that would allow
us to rationally design SEI are still elusive it is simply because
mechanisms cannot be deduced from information on initial and final
states alone. Here, we address this issue in study of the kinetics
of electropolymerization of propylene carbonate (PC) into SEI on metal
electrodes \cite{Osaka_87,An_16}, in conjunction with a fundamental
analysis of the results obtained. Our goal is to gain insight into
the mechanism of SEI generation.

The comparative electropolymerization has been performed through 3
electrolytes given in Table \ref{tab:Elects}. We investigate via
cyclic voltammetry, impedance spectroscopy and chronoamperometry the
role of kinetics in controlling the properties of the SEI generated
from the reduction of propylene carbonate.
\begin{center}
\begin{figure}
\begin{centering}
\includegraphics[width=1\textwidth]{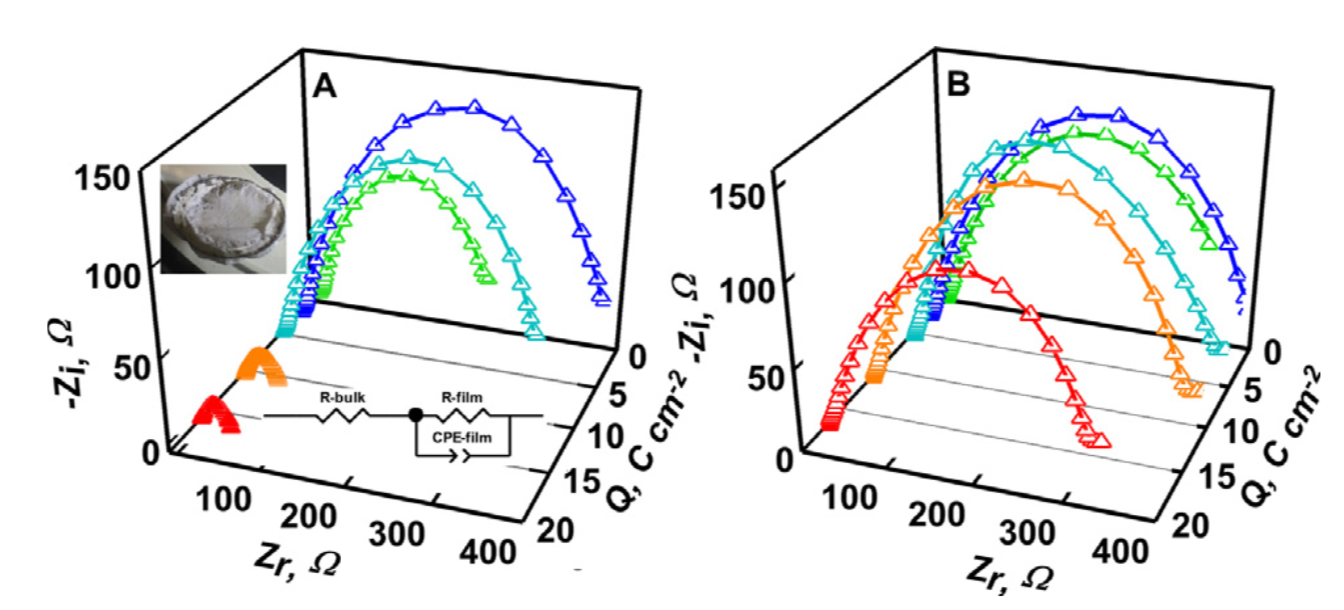}
\par\end{centering}
\caption{{\small{}Nyquist diagrams of $Li|electrolyte|Li$ cells at OCV after
getting charged galvanostatically at $0.05mA/cm^{2}$ for various
charge amounts. }\textbf{\small{}A}{\small{}: with $1M$ $LiClO_{4}$.
}\textbf{\small{}B}{\small{}: with addition of $+0.1M$ $LiF$. A
insets: (1) the equivalent circuit, (2) lithium dendrites that short-circuited
the cell.\label{fig:NyquistSymmetric}\cite{KASMAEE_16}}}
\end{figure}
\par\end{center}

\section{Cyclic Voltammetry}

The CV diagrams for the symmetric cells of electrolytes given in Table
\ref{tab:Elects} with two scanning rates are shown in Figure \ref{fig:CVs1}.
We chose $ClO_{4}^{-}$ because it is a stable, weakly coordinating
anion \cite{Mauro_05} and $F^{-}$ because it is shown to improve
the electropolymerization morphology \cite{Choudhury_16} and PC is
widely used as a base solvent \cite{Nie_15,Profatilova_13}. In figure
\ref{fig:CVs1}\textcolor{red}{A} and \ref{fig:CVs1}\textcolor{red}{B},
peaks between $0.8V$ and $1V$ correspond to PC reduction (PCR hereafter)
\cite{Schaffner_10}. Peaks at $0.3V$ are assigned to the underpotential
deposition (UPD) of $Li^{0}$ on the basis of reported similar peaks
within $0.4\lyxmathsym{\textendash}0.6V$ \cite{Saito_03}, and the
fact that the peak at 1.3 V associated with the anodic stripping of
UPD $Li^{0}$ deposits does not appear following cathodic scans that
were reversed at 0.9 V to avoid $Li^{0}$ deposition.

\section{Electrochemical Impedance Spectroscopy}

The electrical characteristics of the SEI produced can be analyzed
by electrochemical impedance spectroscopy (EIS). The EIS measurements
can be performed at open circuit voltage (OCV). Therefore they only
provide information about the static electrical properties of preformed
SEI. A typical diagrams consist of a single depressed semicircle at
medium and high frequencies, which merges at low frequencies into
a straight line associated with $Li^{+}$ diffusion through SEI layers.
The presence of a single semicircle excludes significant contributions
from multiple SEI layers. Thus SEI properties can be accounted by
a single layer despite their complex, heterogeneous morphology and
chemical composition . In the equivalent circuit shown in the diagram
$R_{Bulk}$ is the sum of ohmic drops across the electrolyte and other
cell components, $R_{SEI}$ and $C_{PE-SEI}$ are the resistance and
capacitance of preformed SEI layers, and $W$ is the impedance arising
from $Li^{+}$ diffusion through SEI layers \cite{Bard_80}.

\begin{table}
\begin{centering}
{\small{}}%
\begin{tabular}{|c|cccccccc|}
\hline 
{\scriptsize{}CV cycle} & \multicolumn{1}{c|}{{\scriptsize{}Electrolyte}} & \multicolumn{1}{c|}{{\scriptsize{}R-Bulk ($\Omega$)}} & \multicolumn{1}{c|}{{\scriptsize{}R-SEI ($\Omega$)}} & \multicolumn{1}{c|}{{\scriptsize{}C-SEI ($10^{5}$$\Omega s^{n}$)}} & \multicolumn{1}{c|}{{\scriptsize{}n}} & \multicolumn{1}{c|}{{\scriptsize{}$\sigma$($10^{4}$$\Omega^{-1}s^{n'}$)}} & \multicolumn{1}{c|}{{\scriptsize{}$n'$}} & {\scriptsize{}$D_{Li^{+}}$($10^{13}$$cm^{2}s^{-1}$)}\tabularnewline
\hline 
\multirow{2}{*}{{\footnotesize{}$1^{st}$}} & {\footnotesize{}I} & {\footnotesize{}4} & {\footnotesize{}543} & {\footnotesize{}1.3} & {\footnotesize{}0.74} & {\footnotesize{}2.1} & {\footnotesize{}0.62} & {\footnotesize{}3.0}\tabularnewline
 & {\footnotesize{}II} & {\footnotesize{}3.1} & {\footnotesize{}459} & {\footnotesize{}1.5} & {\footnotesize{}0.73} & {\footnotesize{}3.6} & {\footnotesize{}0.60} & {\footnotesize{}1.1}\tabularnewline
\cline{1-1} 
\multirow{2}{*}{{\footnotesize{}$5^{th}$}} & {\footnotesize{}I} & {\footnotesize{}4.7} & {\footnotesize{}846} & {\footnotesize{}1} & {\footnotesize{}0.76} & {\footnotesize{}2.1} & {\footnotesize{}0.67} & {\footnotesize{}2.9}\tabularnewline
 & {\footnotesize{}II} & {\footnotesize{}3} & {\footnotesize{}489} & {\footnotesize{}1.6} & {\footnotesize{}0.72} & {\footnotesize{}3.9} & {\footnotesize{}0.62} & {\footnotesize{}0.9}\tabularnewline
\hline 
\end{tabular}{\small\par}
\par\end{centering}
{\small{}\caption{Equivalent circuit parameters from spectra of Figure \ref{fig:CVs1}\textcolor{red}{C}.\label{tab:EquivC}}
}{\small\par}
\end{table}

The implication is that the SEI production is followed by homogeneous
chemical reactions that incorporate substantial amounts of additional
PC into the layers. Constant phase element $Z_{r}$ vs $\omega$ plots:

\begin{equation}
Z_{r}=Z_{0}+\sigma\omega^{-n'}\label{eq:Zr}
\end{equation}

where $Z_{r}$ is the real impedance, $Z_{0}$ is a constant, $\omega$
is frequency, and $r$ and $n'$ are adjustable parameters for Warburg\textquoteright s
impedance associated with $Li^{+}$ diffusion in semi-infinite, homogeneous
SEI layers. The morphology of SEI layers is evidently sensitive to
scan and \emph{PCR} rates. The kinetics of SEI formation. $Li^{+}$
diffusion coefficient, $D^{+}$ in the SEI formed is obtained from

\begin{equation}
D_{Li}^{+}=\frac{R^{2}T^{2}}{2\theta^{2}n^{4}F^{4}[Li^{+}]^{2}\sigma^{2}}\label{eq:D+}
\end{equation}

where $R$ is the gas constant, $T$ is absolute temperature ($298K$),
$\theta$ is electrode area, $n=1$ is the ion charge, $F$ is Faraday\textquoteright s
constant, and $\sigma$ is the slope of $Z_{r}$ vs $\omega^{-n}$
plots. 

The resistance of SEI layers is directly proportional to thickness
l, and electrical resistivity $\rho$:

\begin{equation}
R_{SEI}=\frac{\rho l}{\theta}\label{eq:R-SEI}
\end{equation}

\begin{table}
\begin{centering}
{\small{}}%
\begin{tabular}{|c|cccccccc|}
\hline 
{\scriptsize{}CV cycle} & \multicolumn{1}{c|}{{\scriptsize{}Electrolyte}} & \multicolumn{1}{c|}{{\scriptsize{}R-Bulk ($\Omega$)}} & \multicolumn{1}{c|}{{\scriptsize{}R-SEI ($\Omega$)}} & \multicolumn{1}{c|}{{\scriptsize{}C-SEI ($10^{5}$$\Omega s^{n}$)}} & \multicolumn{1}{c|}{{\scriptsize{}n}} & \multicolumn{1}{c|}{{\scriptsize{}$\sigma$($10^{4}$$\Omega^{-1}s^{n'}$)}} & \multicolumn{1}{c|}{{\scriptsize{}$n'$}} & {\scriptsize{}$D_{Li^{+}}$($10^{13}$$cm^{2}s^{-1}$)}\tabularnewline
\hline 
\multirow{2}{*}{{\footnotesize{}$1^{st}$}} & {\footnotesize{}I} & {\footnotesize{}5.3} & {\footnotesize{}5439} & {\footnotesize{}2.7} & {\footnotesize{}0.79} & {\footnotesize{}3.2} & {\footnotesize{}0.50} & {\footnotesize{}1.3}\tabularnewline
 & {\footnotesize{}II} & {\footnotesize{}4.3} & {\footnotesize{}503} & {\footnotesize{}2.9} & {\footnotesize{}0.79} & {\footnotesize{}3.2} & {\footnotesize{}0.49} & {\footnotesize{}1.2}\tabularnewline
\cline{1-1} 
\multirow{2}{*}{{\footnotesize{}$5^{th}$}} & {\footnotesize{}I} & {\footnotesize{}4.7} & {\footnotesize{}543} & {\footnotesize{}3.1} & {\footnotesize{}0.82} & {\footnotesize{}3.0} & {\footnotesize{}0.46} & {\footnotesize{}1.5}\tabularnewline
 & {\footnotesize{}II} & {\footnotesize{}5.6} & {\footnotesize{}534} & {\footnotesize{}2.8} & {\footnotesize{}0.82} & {\footnotesize{}3.2} & {\footnotesize{}0.46} & {\footnotesize{}1.3}\tabularnewline
\hline 
\end{tabular}{\small\par}
\par\end{centering}
{\small{}\caption{Equivalent circuit parameters from spectra of Figure \ref{fig:CVs1}\textcolor{red}{D}.\label{tab:EquivD}}
}{\small\par}
\end{table}

In contrast, the Nyquist diagrams of SEI grown at $v=0.5\nicefrac{mV}{s}$
(Figure \ref{fig:CVs1}\textcolor{red}{D}) are qualitatively and quantitatively
different from those in Figure \ref{fig:CVs1}\textcolor{red}{C}.
In this case, the SEI produced from electrolytes I and II after the
1st and 5th scans have essentially identical parameters (Table \ref{tab:EquivC}),
which are consistent with electronically insulating and PC-impermeable
SEI layers. The presence of fluoride has a significant effect on the
long-term stability of electropolymerization upon galvanostatic charging
at $0.05\nicefrac{mA}{cm2}$. Note that $PC$ and $Li^{+}$ are simultaneously
reduced during galvanostatic charging. Figures \ref{fig:CVs1}\textcolor{red}{A}
and \ref{fig:CVs1}\textcolor{red}{B} show the evolution of Nyquist
diagrams as functions of circulated charge. Noteworthy is the fact
that the resistance of cells filled with \textbf{II} (containing $F^{-}$)
decreases by only 25\% after the circulation of $Q>17\nicefrac{C}{cm^{2}}$,
whereas the resistance of cells filled with I (without $F^{-}$) already
drops eightfold at $Q>5\nicefrac{C}{cm^{2}}$, as an indication that
$Li^{0}$ dendrites had pierced SEI layers, reached the cathode and
short-circuited the cell. Fluoride additions also enhance the persistence
of electropolymerization.

\section{Chronoamperometry}

PCR at slow scan rates generates PC-impermeable SEI layers (Figures
\ref{fig:CVs1}\textcolor{red}{C} and \ref{fig:CVs1}\textcolor{red}{D})
led us to test the dependence of PCR rates on applied potential by
growing SEI under potentiostatic conditions. CA experiment at 1.0
, 1.1 and 1.7 V (vs $Li^{+}/Li^{0}$)) applied potentials in cells
filled with electrolyte I are shown in Fig. \ref{fig:Chronos}A. Faradaic
currents associated with PCR (i.e., those circulating after the decay
of initial capacitive currents) markedly increase at more negative
overpotentials: $\eta=E-E_{p}$ (PCR rates peak at $E_{p}\sim1.3V$,
Fig. \ref{fig:Chronos}\textcolor{red}{B}), as expected. This confirms
that PCR rates lead to self-healing SEI layers, currents circulating
in the CA at a $\eta=1.7V-1.3V=0.4V$ underpotential vanish after
$\sim5000s$, in contrast with experiments carried at $1.0V$ and
$1.1V$. Past the initial stages where currents are partially due
to the capacitive charging of double layers (and also at $i>50\mu A$,
partially controlled by PC desolvation, cf. Fig. \ref{fig:CVs1}\textcolor{red}{A}
and \ref{fig:CVs1}\textcolor{red}{B}), the slopes of faradaic currents
vs $(time)^{-\nicefrac{1}{2}}$:

\begin{equation}
i=\frac{nF\theta[PC]\sqrt{D_{PC}}}{\sqrt{\pi t}}\label{eq:i}
\end{equation}

\begin{table}
\begin{centering}
{\small{}}%
\begin{tabular}{|ccccccccc|}
\hline 
\multicolumn{1}{|c|}{{\scriptsize{}V (vs $Li^{+}/Li^{0}$)}} & \multicolumn{1}{c|}{{\scriptsize{}Electrolyte}} & \multicolumn{1}{c|}{{\scriptsize{}$R_{Bulk}$($\Omega$)}} & \multicolumn{1}{c|}{{\scriptsize{}$R_{SEI}$ ($\Omega$)}} & \multicolumn{1}{c|}{{\scriptsize{}$C_{SEI}$($10^{5}$$\Omega s^{n}$)}} & \multicolumn{1}{c|}{{\scriptsize{}n}} & \multicolumn{1}{c|}{{\scriptsize{}$\sigma$($10^{4}$$\Omega^{-1}s^{n'}$)}} & \multicolumn{1}{c|}{{\scriptsize{}$n'$}} & {\scriptsize{}$D_{Li^{+}}$($10^{13}$$cm^{2}s^{-1}$)}\tabularnewline
\hline 
\multirow{2}{*}{{\footnotesize{}1.0}} & {\footnotesize{}1st} & {\footnotesize{}30} & {\footnotesize{}800} & {\footnotesize{}0.89} & {\footnotesize{}0.77} & {\footnotesize{}0.35} & {\footnotesize{}0.85} & {\footnotesize{}110}\tabularnewline
 & {\footnotesize{}2nd} & {\footnotesize{}35} & {\footnotesize{}1641} & {\footnotesize{}0.77} & {\footnotesize{}0.72} & {\footnotesize{}0.34} & {\footnotesize{}0.90} & {\footnotesize{}120}\tabularnewline
\hline 
\multirow{2}{*}{{\footnotesize{}1.1}} & {\footnotesize{}1st} & {\footnotesize{}14} & {\footnotesize{}920} & {\footnotesize{}0.95} & {\footnotesize{}0.73} & {\footnotesize{}0.79} & {\footnotesize{}0.78} & {\footnotesize{}21}\tabularnewline
 & {\footnotesize{}2nd} & {\footnotesize{}15} & {\footnotesize{}1336} & {\footnotesize{}0.72} & {\footnotesize{}0.73} & {\footnotesize{}0.71} & {\footnotesize{}0.77} & {\footnotesize{}26}\tabularnewline
\hline 
\multirow{2}{*}{{\footnotesize{}1.7}} & {\footnotesize{}1st} & {\footnotesize{}16} & {\footnotesize{}616} & {\footnotesize{}1.8} & {\footnotesize{}0.79} & {\footnotesize{}1.4} & {\footnotesize{}0.73} & {\footnotesize{}7.3}\tabularnewline
 & {\footnotesize{}2nd} & {\footnotesize{}16} & {\footnotesize{}565} & {\footnotesize{}1.6} & {\footnotesize{}0.80} & {\footnotesize{}2.0} & {\footnotesize{}0.71} & {\footnotesize{}3.4}\tabularnewline
\hline 
\end{tabular}{\small\par}
\par\end{centering}
{\small{}\caption{Equivalent circuit parameters from spectra of Figure \ref{fig:CVs1}\textcolor{red}{A}.
\ref{fig:CVs1}\textcolor{red}{C}.\label{tab:Equiv2}}
}{\small\par}
\end{table}

lead to vastly different PC diffusion coefficients in SEI layers grown
at 1.0 V: $D_{PC}(1.0V)=8.3\times10^{-14}cm^{2}s^{-1}$ vs. those
grown at 1.7 V: $D_{PC}(1.7V)=7.7\times10^{-17}cm^{2}s^{-1}$, which
are compatible with the $D_{PC}\sim10^{-12}$ to $10^{-16}cm^{2}s^{-1}$
values reported in porous and compact SEI layers, respectively \cite{Guan_15}.
Note that Eq. \ref{eq:i} for PC diffusion through a growing solid
SEI layer is the analogue of Cottrell\textquoteright s equation for
ion diffusion through a widening, solvent-filled double layer. In
both cases layer thicknesses increase with $t^{\nicefrac{1}{2}}$,
and the corresponding current densities decrease with $t^{-\nicefrac{1}{2}}$
\cite{Bard_80}. Most remarkably, $D_{PC}(1.7V)$ is $\sim1100$ times
smaller than $D_{PC}(1.0V)$ through SEI layers that were seeded by
a small fraction of the charge: $Q_{1.7V}/Q_{1.0V}=0.04$ ($Q=\int Idt$)
\begin{center}
\begin{figure}
\begin{centering}
\includegraphics[width=1\textwidth]{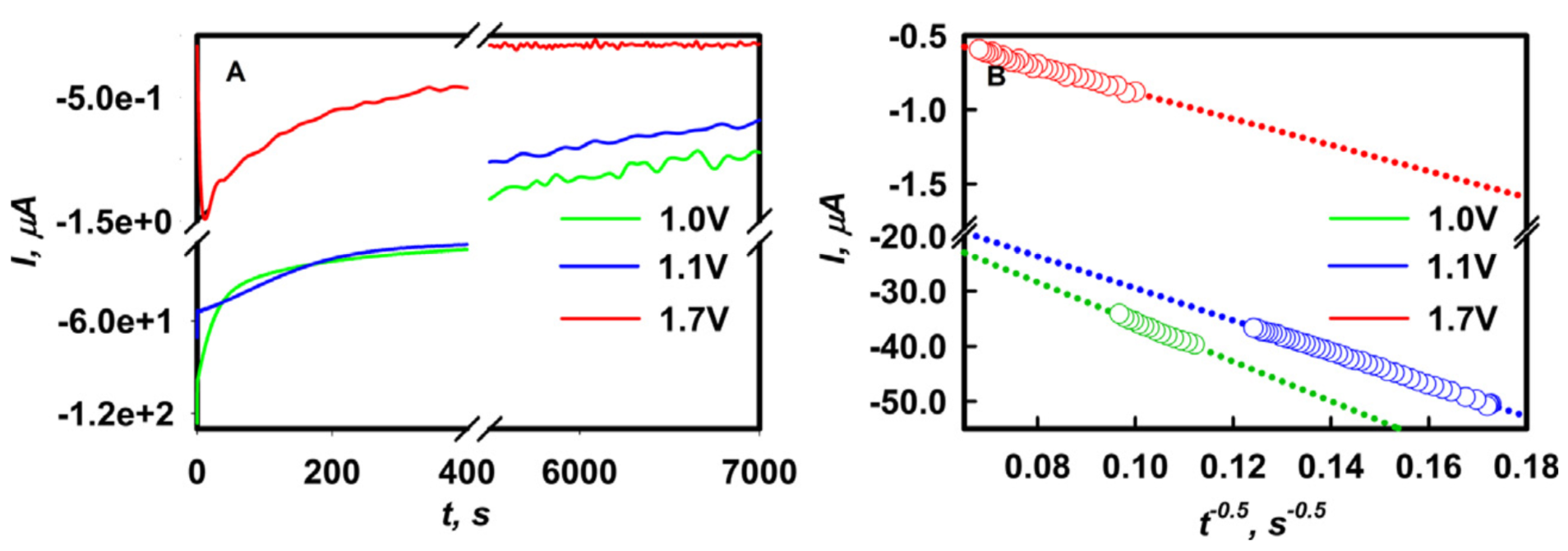}
\par\end{centering}
\caption{\textbf{\small{}A}{\small{}: Chronoamperograms in Cu|electrolyte|Li
cells filled with electrolyte }\textbf{\small{}I}{\small{} under $1.0V$,
$1.1V$ and $1.7V$ applied voltages (vs 1M $Li^{+}/Li^{0}$ in PC).
}\textbf{\small{}B}{\small{}: Cottrell current vs. $t^{-0.5}$ plots
(Eq. \ref{eq:i}).\label{fig:Chronos}\cite{KASMAEE_16}}}
\end{figure}
\par\end{center}

The relevant electrochemical characteristics of the SEI layers grown
potentiostatically were probed by EIS and CV experiments. Nyquist
diagrams of the SEI layers produced in successive chronoamperometry
experiments under 1.0, 1.1 and 1.7 applied potentials are shown in
Fig. \ref{fig:Chronos}\textcolor{red}{A}. The parameters derived
from their analysis are compiled in Table \ref{tab:Equiv2}. It is
apparent that the SEI produced in the first 1.7 V potentiostatic experiment
does not grow upon further charging, in contrast with those produced
at 1.0 and 1.1 V. This conclusion is corroborated by CV scans (Fig.
\ref{fig:Chronos}\textcolor{red}{B}).

Inspection of Table \ref{tab:Equiv2} reveals that:
\begin{enumerate}
\item $R_{SEI}$ of the layers grown in initial cycles is comparable values
despite the fact that the amount of PC reduced from electrolyte. $R_{SEI}$
remains nearly constant in 1.7 V experiments but increases by a factor
of 2 in the second CA at 1.0 V.
\item $C_{SEI}$ at 1.7 V is about $\times2$ times larger than those at
1.0 and 1.1 V suggesting (since ${\displaystyle C\propto\frac{1}{thickness}}$)
that they are about half as thick.
\item $Li^{+}$ diffusion, with $n_{0}>0.7>0.5$ , is anomalous in all cases.
Noteworthy is that $D^{+}$ for SEI layers produced at 1.7 V is comparable
to the $D^{+}$ values in the SEI obtained in potentiodynamic CV experiments
(see Tables \ref{tab:EquivC} and \ref{tab:EquivD}), but much smaller
than $D^{+}$ in SEI layers produced at 1.0 Li and 1.1 V, as evidence
that SEI morphology is a sensitive function of applied potentials.
\end{enumerate}
Summing up, the above findings are consistent with:
\begin{enumerate}
\item SEI layers that incorporate PC molecules in larger numbers than those
undergoing reduction at the electrode surface, i.e., SEI are essentially
polymer materials \cite{Tavassol_12}.
\item SEI properties strongly depend on the kinetics of the generation process.
\cite{Koh_13}
\begin{figure}
\begin{centering}
\includegraphics[width=1\textwidth]{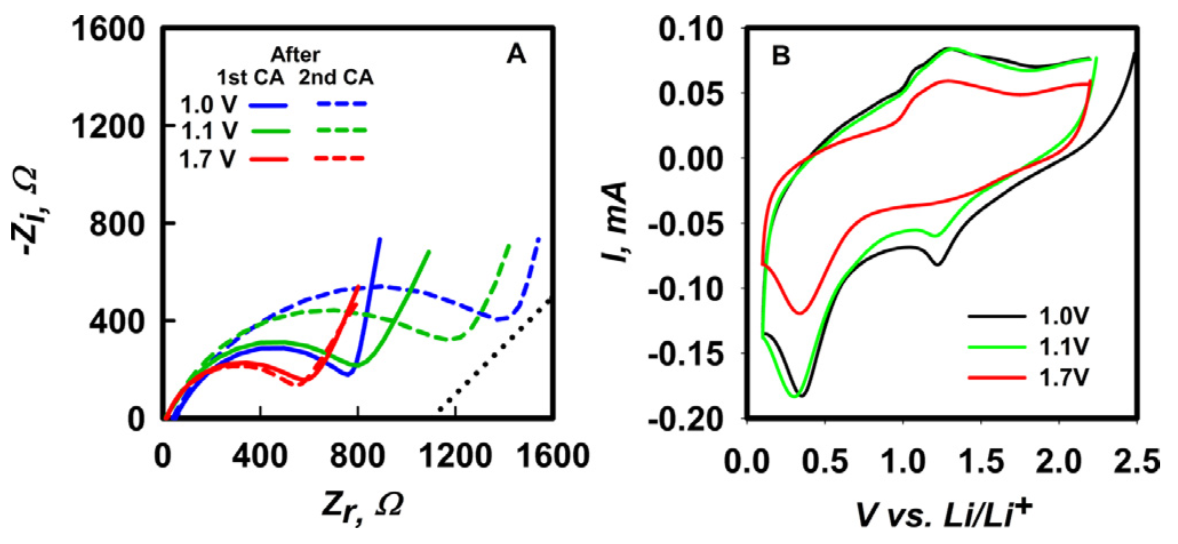}
\par\end{centering}
\caption{\textbf{\small{}A}{\small{}: Nyquist diagrams at open circuit voltage
of Cu|electrolyte|Li cells filled with electrolyte I after the first
and second chronoamperometries at 1.0 V, 1.1 V and 1.7 V applied voltages
vs. 1 M $Li^{+}/Li^{0}$ in PC. Dotted line: Warburg\textquoteright s
$n'=0.5$ slope as a reference. }\textbf{\small{}B}{\small{}: CV at
$v=5mV/s$ in cells with electrolyte I after being charged potentiostatically
at 1.0 V, 1.1 V and 1.7 V for 2 h.\label{fig:Cvs2}\cite{KASMAEE_16}}}
\end{figure}
\end{enumerate}
Since SEI behave as polymeric materials, our findings suggest that
the potential impact of experimental conditions on their properties
should be evaluated on the basis of polymer science concepts \cite{Flory_53,Cosnier_11}.
The radical chain PC electropolymerization into polymer units whose
complexity increases at lower initiation rates. We show that slow
initiation rates via one-electron PC reduction at underpotentials
consistently yields compact, electronically insulating, Li+-conducting,
PC- impermeable SEI films.What to expect for SEI generated in a polymerization
process initiated by PC reduction, reaction eq:PC: \cite{Seo_14,Gachot_08}

\begin{equation}
PC+e^{-}\rightarrow PC^{.-}\label{eq:PC}
\end{equation}

Following previous reports \cite{Tavassol_12,Shkrob_13}, PC is deemed
to open its ring into an alkoxycarbonyl radical followed by decomposition
into $CO$ or $CO_{2}$, plus simpler radical anions, $X^{.-}$, which
initiates radical chain-growth polymerizations propagated by reactions
\ref{eq:X}:

\begin{equation}
X^{.-}+PC\rightarrow X-(PC){}^{.-}\rightarrow\rightarrow X-(PC)_{n}^{.-}\label{eq:X}
\end{equation}

and terminated via bimolecular radical recombination, reaction \ref{eq:NM}:
\cite{Soto_15}

\begin{equation}
X-(PC)_{n}^{-}+X-(PC)_{m}^{-}\rightarrow X_{2}-(PC)_{n+m}^{2-}\label{eq:NM}
\end{equation}

SEI permeability, ionic and electronic conductivity, solubility and
mechanical properties are essentially determined by the degree of
solvent polymerization $\lambda$, i.e., by the number of monomers
incorporated into polymer units \cite{Shkrob_13}. $\lambda$ is controlled
by the competition between radical propagation (Reaction \ref{eq:X})
vs. radical termination (Reaction \ref{eq:NM}). Thus we arrive at
Reaction \ref{eq:lambda}:

\begin{equation}
\lambda=\frac{k_{2}[PC^{.-}][PC]}{2k_{3}[PC^{.-}]^{2}}=\frac{k_{2}[PC]}{2k_{3}[PC^{.-}]}\label{eq:lambda}
\end{equation}

where $k_{2},k_{3}$are bimolecular reaction rate constants. Because
initiation rates $r_{i}$ and termination rates balance at steady
state, we have:

\begin{equation}
r_{i}=2k_{3}[PC^{.-}]^{2}\Rightarrow[PC^{.-}]=\left(\frac{r_{i}}{2k_{3}}\right)^{0.5}\label{eq:ri}
\end{equation}

Therefore we arrive at Reaction \ref{eq:lambda2}:

\begin{equation}
\lambda=\frac{k_{2}[PC]}{\sqrt{2k_{3}r_{i}}}\label{eq:lambda2}
\end{equation}

which predicts that the degree of polymerization should be directly
proportional to PC concentration at the front of advancing radical
chains, and inversely proportional to (initiation rates: $r_{i}=r_{1}$)$^{\frac{1}{2}}$.
The fundamental $\lambda\propto r^{{\displaystyle \nicefrac{1}{2}}}$relationship
for a radical chain polymerization should apply whether $Li^{+}$
are present in the SEI, as in the present case, or not.

Thus, our CA experiments at 1.7 V are deemed to produce functional
SEI layers because the low current densities exclusively associated
with PC reduction provide the slow initiation rates required to generate
long polymerization chains. Furthermore, as a result, the overall
slow polymerization process they bring about may not be limited by
the availability of the free PC monomers released from the slow desolvation
of $Li(PC)_{n}^{+}$. The very low value of the PC diffusion coefficient
$D_{PC}(1.7V)=7.7\times10^{-17}cm^{2}s^{-1}$ determined in the SEI
generated at underpotential is clearly consistent with transport through
a compact material comprising few, long and possibly linked or intertwined
polymer chains \cite{Guan_15}. From this perspective, the PC reduction
rates at the $\sim1V$ overpotentials prevailing under conventional
LMB charging conditions, where the full voltage required to plate
the anode is applied from the onset, may not be ideal because they
are likely to generate short, disjoint polymer domains rather than
compact, interconnected polymer films extending over the electrode
surface. We believe that our results and analysis provide new insights
into the outstanding questions formulated in a recent review on the
subject: \textquoteleft how does SEI form?\textquoteright{} and \textquoteleft what
parameters control SEI properties?\textquoteright{} \cite{Gauthier_15}.

\section{Experimental}

Experiments were performed in two types of electrochemical cells filled
with electrolyte solutions I, II and III of three different compositions
(Table 1). Studies on SEI layers were carried out in $Cu|electrolyte|Li$
coin cells whereas the deposition of $Li^{0}$ films was investigated
in $Li|electrolyte|Li$ coin cells. Round disk electrodes ($A=1.6cm^{2}$)
were punched from $Li^{0}$ foil (Aldrich, 99.9\%, $0.38mm$ thick)
that had been polished by scraping with a blade and rinsed with dimethyl
carbonate. Electrodes were mounted on a transparent poly-methyl methacrylate
separator that kept them $L=3.175mm$ apart. All operations were carried
out in a glove box sparged with argon. Chronoamperometry (CA), electrochemical
impedance spectroscopy (EIS) and cyclic voltammetry (CV) measurements
were made with a Bio-Logic VSP potentiostat. Galvanostatic experiments
were performed with an ARBIN BT2000 battery tester. EIS experiments
($5mV$ modulation signal amplitude) covered the $100mHz$ to $1MHz$
frequency range. Impedance data were analyzed using \emph{Zview} software.
All reported potentials are relative to $Li^{+}/Li^{0}$ under working
conditions. $Li^{0}$ and $Cu^{0}$ foils (Sigma-Aldrich) were used
as-received. Lithium per- chlorate ($LiClO_{4}$, Aldrich, battery
grade, $99.99\%$) and lithium fluoride ($LiF$, Aldrich, $99.99\%$
trace metal basis) were dried at $90C$ under vacuum for $24h$ and
dissolved in propylene carbonate (PC) (Aldrich, 99.7\% anhydrous).
Further details can be found in previous publications from our laboratory.
\cite{ARYANFAR_14,ARYANFAR_15_2,ARYANFAR_15,ARYANFAR_14_2}

\section*{Acknowledgement}

Authors would like to thank Prof. William Goddard \footnote{Chemistry and Chemical Engineering, California Institute of Technology,
Pasadena, CA 91125;} and Prof. Jaime Marian\footnote{Material Science and Engineering, University of California, Los Angeles,
CA 90095;} for their insightful comments in various instances.

\bibliographystyle{unsrt}
\bibliography{/Users/aryanfar/Dropbox/PAPERS/Refs}

\end{document}